\documentstyle[preprint,eqsecnum,aps]{revtex}

\begin{document}
\preprint{CGPG-94/4-5, gr-qc/9404053}
\title{Quantum Observables and Recollapsing Dynamics}
\author{Donald Marolf\cite{Marolf}}
\address{Department of Physics, University of California,
Santa Barbara, California 93106}
\date{revised: September 1994}
\maketitle

\begin{abstract}
Within a simple quantization scheme,
observables for a large class of finite dimensional
time reparametrization
invariant systems may be constructed
by integration over the manifold of time labels.  This
procedure is shown to produce a complete set
of densely defined operators on a physical
Hilbert space for which an inner product is identified and to provide
reasonable results for simple test cases.  Furthermore,
many of these
observables have a clear interpretation in the classical limit and we use
this to demonstrate that, for a class of minisuperspace models including LRS
Bianchi IX and the Kantowski-Sachs model this
quantization
agrees with classical physics in predicting that such spacetimes
recollapse.

\end{abstract}
\pacs{ICP: 0460}

\section{Introduction}
\label{intro}
A generalization of Kucha\v{r}'s ``Hilbert space problem" \cite{Kuchar}
provides one of the
greatest obstructions
to developing a quantum theory of gravity.  At issue are
the questions ``What are the physical (quantum) observables?" and ``On what
Hilbert space (if any) do they act?"  This includes the familiar question
``What is the inner product on the space of physical states?"  Here,
we combine the idea emphasized by DeWitt \cite{Bryce}
of constructing ``local" diffeomorphism invariants
by integrating singular densities
over a manifold with the quantization of constrained systems
described by Dirac
\cite{PAMD} and find that
we are able to construct gauge invariant
operators which
(at least in the classical limit) have a clear
interpretation and which are densely
defined on a Hilbert space of physical states with a known inner product.
This approach has the advantage that it is not necessary to identify
an ``internal time function" \cite{Kuchar}, that it may be applied to a
large class
of minisuperspace models -- which includes the LRS Bianchi IX and
Kantowski-Sachs cosmologies --
and that for the models just mentioned it predicts that the
corresponding spacetimes recollapse, in accord with the classical result.
Note that it is
highly nontrivial for a quantization scheme to
satisfy this last condition and, in particular, that
procedures based on
a Klein-Gordon type inner product instead describe the universe
as expanding ``forever" \cite{KG}.

Let us consider for a moment the more general situation of a
diffeomorphism invariant system on a spacetime manifold $M$.
Some interesting common examples are general relativity,
topological field theories, parametrized Newtonian systems, and
homogeneous
cosmological models (such as the Bianchi models) for which $M$ is just
the manifold ${\bf R}$ of time labels.
Solving the equations of motion for such systems
introduces a number of arbitrary functions corresponding to the freedom
to choose coordinates on $M$.  Physical information is
contained in gauge invariant quantities constructed from such
solutions (that is, quantities which do not depend on
these arbitrary functions) so that, classically, we face the issue
of constructing such objects and extracting useful information.
We will refer to such gauge invariants as ${\it observables}$.

As has been pointed out a number of times (see, e.g. \cite{Bryce}),
we {\it do} in fact know how
to construct a great number of observables
for classical diffeomorphism invariant theories.  Given any density
$\omega$ on an n-manifold $M$, $\int_M \omega$ is such an observable.
Of course, these observables are not of the type with
which we are familiar from,
say, scalar field theory and they may be highly non-local.
This is the case for simple examples such as $\int_M \sqrt{-g} R$ in
Einstein gravity.  Also, if the
density $\omega$ is not chosen carefully, such an integral may converge
only when evaluated for very special solutions of the dynamics.

However, as described in \cite{Bryce}, some observables of this type {\it can}
be easily interpreted.  If the density $\omega$ is distributional,
the resulting gauge invariant may be effectively local on $M$.
A simple example of such an observable is the value of some
scalar quantity at a point specified by an ``intrinsic coordinate system."
For gravity coupled to a set of scalar fields $\phi^k$, $k \in
\{0,1,2,3\}$, the corresponding
integral might be
\begin{equation}
\label{int coord}
\int_M  R \prod_{i= 0}^3 \delta(\phi^i ) \Big| {{\partial \phi^k}
\over {\partial x^{\mu}}} \Big| d^4x
\end{equation}
where $R$ is the curvature scalar and
$\Big| {{\partial \phi^k}\over {\partial x^{\mu}}} \Big|$ is a Jacobian.
This object defines a  well-behaved function on any part of the classical
space of solutions for which the fields $\phi^k$ vanish simultaneously
only at a finite
number of spacetime points and for which the metric is smooth
whenever $\phi^k = 0$.

We would like to use similar ideas to define and
discuss observables of this type for {\it quantum} theories with
reparametrization or diffeomorphism invariance.  Two standard criticisms
of observables of this type in the classical setting are 1) that they
will in general be well-defined (and finite)
only on a part of the classical solution space
and 2) that, as they involve integration over {\it time}, they are
likely to be of
little use unless the theory can be solved completely (so
that the integrand may be written explicitly).  Nevertheless,
we will see that quantum analogues of these observables
exist as densely defined operators on a space of physical
states for interesting models and that certain properties
of the resulting operators can be studied even though the models may not
be exactly solvable.  In particular, such observables can be used to
show that, in correspondence with the classical result,
our quantization of appropriate minisuperspace models
describes spacetimes whose homogeneous slices
expand to some maximal size and then recollapse.

The strategy is as follows.  First, section \ref{class} discusses the
classical form of the observables of interest.  This will
provide insight into the definitions and arguments used in the quantum
treatment.
We then give the general quantum formalism for the operators
and ths physical Hilbert space in \ref{Dirac} and apply it to
two test cases in \ref{se}.  After verifying that
appropriate results are produced for these simple cases,
we turn to the cosmological models of interest in \ref{sep}.
In each application, the
integrals that define our observables will be shown to converge
and we find a ``complete set" of observables that are densely defined
on the physical Hilbert space and symmetric with respect to a physical
inner product.  In addition, \ref{sep}
shows that when applied to ``separable semi-bound" cosmological models,
this quantization describes spacetimes that recollapse.  We conclude
with a short discussion in section \ref{diss}.

\section{Observables through Integration}
\label{class}

Section \ref{Dirac} will address quantization of reparametrization
invariant models and will focus on the properties of certain
observables.  In order to provide an intuitive understanding of the
definition and use of these observables, we now
discuss
their classical counterparts.  Thus, this section can be regarded as
essentially heuristic background for the more rigorous results of
\ref{Dirac}, \ref{se}, and \ref{sep}.

We are interested in systems that are time reparametrization invariant;
that is, systems defined on some $d+1$ dimensional spacetime $M$ for
which there is a gauge symmetry that takes $f(t)$ to any
$f(T(t))$ whenever $dT/dt >0$ and $f$ is a scalar on $M$.  We
concentrate on the case $d=0$ and on observables
like \ref{int coord}. However, because canonical techniques will be
used, it is better to consider invariants of the form

\begin{equation}
\label{inv2}
[a]_{q = \tau} = \int_{-\infty}^{\infty} dt \ {{dq} \over {dt}}
\delta(q(t) - \tau) a(t)
\end{equation}
which may be regarded (when the integral converges) as either
functions on the space ${\cal S}$ of classical solutions or,
through the evolution map that constructs a solution
from a piece of initial data specified by a point on the constraint
surface, on the
part of the corresponding phase space in which any constraints
are satisfied.  When $q(t)$ takes the value $\tau$ once along a solution, this
object gives the value of $a(t)$ when $q(t) =
\tau$ up to a sign.  Similar observables were used in
\cite{tril,Carlo,Lee}.

The advantage of \ref{inv2} is that it is invariant not just
under reparametrizations ($t \rightarrow T(t)$ with $dT/dt >0$), but
also under pull backs of the manifold of time labels through
maps of degree 1 ($t \rightarrow
T(t)$ for any smooth $T(t)$ of degree 1).  This large class of transformations
typically become gauge symmetries as well when a time reparametrization
invariant system is written in Hamiltonian form.

In fact, if the canonical description is in terms of
some phase space coordinates $z_i(t)$ at each $t$, a (constrained)
Hamiltonian $h$
and a lapse function $N(t)$, it typically has
\begin{eqnarray}
\label{Hgts}
\delta z_i(t) &=& \epsilon (t) \ \{h, z_i\} = - {{\epsilon (t)} \over {N(t)}}
{{\partial z_i(t)} \over {\partial t}} \cr
\delta N(t) &=& - {{\partial} \over {\partial t}} \epsilon (t)
\end{eqnarray}
as a gauge symmetry. Here
$\{,\}$ is the canonical Poisson Bracket and $\epsilon (t)$ is the gauge
parameter.

Because canonical methods will be used, we invoke an
abuse of language and refer to transformations of the form \ref{Hgts}
as ``reparametrizations" throughout this work.  Reparametrizations of
the first type ({\it diffeomorphisms} of the $t$-label set) will be referred
to as ``true reparametrizations."  Note that the infinitesimal
form of a true reparametrization $t \rightarrow t + \delta t$ is
identical to that of a general reparametrization with parameter
$\epsilon = N \delta t$.
Because its integrand is a one-form,
expression \ref{inv2} is in fact invariant under the more general
reparametrizations (\ref{Hgts}).

Along a solution on which $q=\tau$ more than once or not at all,
the question ``What is the value of $a(t)$ when
$q(t) = \tau$" may be ill-defined.
However, expression \ref{inv2} provides a
natural generalization of the ``answer."
It adds the values of $a(t)$ at each $t$ such that
$q(t) = \tau$ and ${{dq} \over {dt}} > 0$ and subtracts those
values of $a(t)$ for which $q(t) = \tau$ and ${{dq} \over {dt}} < 0$.
Of course, for some solutions $s$ the integral in \ref{inv2} may not converge,
such as when $q(t)$ takes the value $\tau$ an infinite number of times along
$s$, and on such solutions $[a]^{(2)}_{q = \tau}$ is undefined.
However, this will
not be a problem for the cases we consider.

Note that if $q=\tau$ at several points along a solution,
a calculation of $[a]_{q=\tau}$ involves cancellations between those
points with $q=\tau$ at which $dq/dt >0$ and $dq/dt < 0$.  To
obtain an average without such cancellation while retaining
invariance under \ref{Hgts}, we might consider $[b]_{q=\tau}$ for
$b = a \ {\rm{sign}} \Bigl(
{1 \over N} {{dq} \over {dt}} \Bigr)$ where ${\rm{sign}}(x) = \theta(x) -
\theta(-x)$.
In particular, $[{\rm{sign}}\Bigl( {1 \over N} {{dq} \over {dt}}
\Bigr)]_{q = \tau}$
is the number of times for which $q(t) = \tau$ and $N(t) > 0$ minus
the number of times for which $q(t) = \tau$ and $N(t) < 0$.  In a
``proper parametrization" for which $N(t) > 0$ and proper time always flows
forward, this is just the number of times for which $q(t) = \tau$.

Thus, if we are considering a reparametrization invariant
minisuperspace model and $q$ is some measure of the size of
the universe, the statement that
\begin{equation}
\label{class recol}
\lim_{\tau \rightarrow \infty} \Bigl[{\rm{sign}} \Bigl( {1 \over N}
{{dq} \over {dt}} \Bigr) \Bigr]
_{q = \tau} = 0
\end{equation}
pointwise on ${\cal S}$ (or the on the constraint surface) is
equivalent to the statement that such models always
recollapse; i.e., that any given solution $s$ does not reach arbitrarily large
values of $q$.  For such a model, we in fact have
\begin{equation}
\label{all a}
\lim_{\tau \rightarrow \infty} [a]_{q = \tau} = 0
\end{equation}
for any $a = a(z_i)$.
In \ref{sep}, we will deduce an analogous property
of corresponding quantum operators for certain
minisuperspace models.  We interpret this as a demonstration that our
quantization of these models describes recollapsing cosmologies.

\section{General Dirac Methods}
\label{Dirac}

We now use the heuristic ideas embodied in \ref{inv2} to quantize time
reparametrization invariant systems.  This section gives the general
setting and formalism to be used while important examples are described in
the next two sections.  When the quantum version of the integral
\ref{inv2} converges, we will define the $[A]_{Q=\tau}$
through integration and show in \ref{QO} that they are
observables in the canonical sense; that is, that they commute with
the Hamiltonian.  We introduce a space of physical states as in
\cite{PAMD} and find
a physical inner product in \ref{in prod}
with respect to which our operators are symmetric.
The results of this section assume convergence
of the integrals, which will be separately discussed
for each of the
examples of \ref{se} and \ref{sep}.  However, results for
systems outside the scope of \ref{se} and \ref{sep} should be considered
as formal.

\subsection{The Setting}
\label{co}

Let us suppose that we wish to quantize a
time reparametrization invariant system
described by a
phase space which is a cotangent bundle $T^*{\cal Q}$ over
a configuration space ${\cal Q}$.  We will also assume that the
dynamics is expressed through  a Hamiltonian $h$ and lapse
function $N$.  Because time reparametrization is to be a
gauge symmetry of this system, the Hamiltonian is classically constrained
to vanish: $h = 0$.

If we momentarily ignore the reparametrization invariance and the
constraint, this system has a straightforward quantization.
States of the
system live in the Hilbert space $L^2({\cal Q})$ and
the Hamiltonian $h$ becomes a Hermitian
operator $H$.  In general, we will use capital letters for
quantum operators and lower case letters for their eigenvalues and
classical counterparts.
$H$ generates a unitary (Newtonian) time evolution
and may be used to define a set of
Heisenberg picture operators corresponding to the functions and vector
fields on ${\cal Q}$ in the usual way.
In order for $H$ to be self-adjoint, the
$L^2({\cal Q})$ functions may be required to
satisfy additional boundary conditions.

Thus, our only tasks are to impose the constraint and to implement the
time reparametrization invariance.  For the second of these,
we introduce
a lapse operator $N(t)$ and an evolution in a {\it parameter} time $t$
instead of the Newtonian time evolution described above.
We proceed with the following algebraic approach:

For each smooth function $q$ and each vector field $p$ on
${\cal Q}$, introduce one parameter families $Q(t)$ and $P(t)$
of elements in a noncommutative algebra.  Recall that such $q$ and $p$
form an overcomplete set of functions on $T^*{\cal Q}$ and that the
classical Poisson Brackets are summarized by
\begin{eqnarray}
& \{q_1, q_2 \} = 0 \cr
& \{q_1, p_1 \} = {\cal L}_{p_1}q_1 \equiv q'(q_1,p_1) \cr
& \{p_1, p_2 \} = \{p_1,p_2\}_{\cal L}  \equiv p'(p_1,p_2)
\end{eqnarray}
where $q_1,q_2$ are smooth functions on ${\cal Q}$, $p_1,p_2$ are smooth
vector fields on ${\cal Q}$, ${\cal L}_{p_1}$ denotes the Lie derivative
along $p_1$, $\{,\}_{\cal L}$ is the Lie Bracket of vector fields, and
the above expressions define the function $q'$ and the vector field
$p'$ in terms of $q_1$, $p_1$, and $p_2$.
We therefore impose the relations:
\begin{eqnarray}
\label{quant alg}
& [Q_1(t), Q_2(t)] = 0 \cr
& [Q_1(t), P_1(t)] = Q'(q_1,p_1) \cr
& [P_1(t), P_2(t)] = P'(p_1,p_2)
\end{eqnarray}
on our algebra.  We also introduce the family $N(t)$, each
member of which commutes with
every element in the algebra.  Further, we require
\begin{eqnarray}
\label{dyn}
& -i {{\partial} \over {\partial t}} Q(t) = N(t) [H(t), Q(t)] \cr
& -i {{\partial} \over {\partial t}} P(t) = N(t) [H(t), P(t)]
\end{eqnarray}
where $H(t)$ is some symmetric factor ordering of the
classical expression for $H$ in terms of the $q$'s and $p$'s
with these objects replaced by the algebraic elements $Q(t)$ and $P(t)$.
While this choice of algebra is not unique (see \cite{hisb,gp}) we
will not consider other possibilities here.

The presence of the lapse makes the dynamics reparametrization
invariant.  That is, the above equations are unchanged by the
transformations
\begin{eqnarray}
\label{gts}
\delta Q(t) &=& -i \epsilon(t) \ [H(t), Q(t)] = - {{\epsilon
(t)} \over {N(t)}}
{{\partial Q(t)} \over {\partial t}} \cr
\delta P(t) &=& -i \epsilon (t) \ [H(t), P(t)] = -
{{\epsilon (t)} \over {N(t)}}
{{\partial P(t)} \over {\partial t}} \cr
\delta N(t) &=& - {{\partial} \over {\partial t}} \epsilon (t)
\end{eqnarray}
We may thus consider \ref{dyn} as simply
a Heisenberg picture version of the Schr\"odinger equation imposed in
\cite{PAMD}.

As in unconstrained quantum mechanics, we now search for
irreducible Hilbert space
representations ${\cal H}$ of $\ref{quant alg}$. For fixed $t$, they can
be constructed in the usual way through ``coordinate
representations" carried by square integrable
functions on ${\cal Q}$, on which
the $Q(t)$ act by multiplication by $q$ and the $P(t)$ generate
diffeomorphisms along the integral curves of $p$.  Since we require
$H$ to be Hermitian, the
dynamics \ref{dyn} identifies representations of operators at
the times $t$ and $t'$ through
\begin{eqnarray}
& Q(t) = \exp \Bigl( i H \int_{t'}^t dt N(t) \Bigr) Q(t')
\exp \Bigl( -i H \int_{t'}^t dt N(t) \Bigr)  \cr
& P(t) = \exp \Bigl( i H \int_{t'}^t dt N(t) \Bigr) P(t')
\exp \Bigl( -i H \int_{t'}^t dt N(t) \Bigr)
\end{eqnarray}
where, as usual $H =  H(t)$ is actually time independent.  Since $N(t)$
commutes with every operator, we have
$N(t) = n(t) \openone$ for some real-valued $n(t)$ in any irreducible
representation for which $N(t)$ is Hermitian.  From now on, we will
assume that $n(t)$ satisfies
\begin{equation}
\label{n cond 1}
\lim_{t' \rightarrow \pm \infty} \int_0^{t'} n(t)dt = \pm \infty.
\end{equation}
All such smooth choices of $n(t)$ are equivalent and are related
to each other by gauge transformations.
Note that, given a choice of $n(t)$, this representation is unique to
the same extent as in usual quantum mechanics. In particular,
if the phase space is $T^*{\bf R}^n$ with the standard symplectic
structure, it is unique by the Stone-von Neumann theorem if weak
continuity is assumed.

In order to enforce the constraint in our quantum theory,
we will say that ``physical" states $|\psi\rangle$ lie in the physical
subspace ${\cal H}_{phys}$ determined (as in \cite{PAMD}) by
\begin{equation}
H | \psi \rangle = 0
\end{equation}
In general, such states will only be delta-function
normalizable. That is, such states will not lie in ${\cal
H}_{aux}$ at all, but in the dual to the nuclear space associated with
the spectrum of $H$.
We will deal with this feature later.

\subsection{The Quantum Observables}
\label{QO}

Our task will be to construct
{\it observables}; that is, operators which are invariant under
\ref{gts}.  We will again use operators that correspond to the
classical objects $[a]_{q=\tau}$ defined through:
\begin{equation}
\label{qi}
[A]_{Q = \tau} = \int_{-\infty}^{\infty} dt \ {{dQ} \over {dt}}
\delta(Q(t) - \tau) A(t)
\end{equation}
where, for the moment, we will not worry too much about the
factor ordering of the integrand or the definition of the
delta-function.
To show that $[A]_{Q = \tau}$ commutes with $H$, we will prove the
more general result that time reparametrization invariant operators
of the form
\begin{equation}
\label{form}
\Omega = \int \omega(t) dt
\end{equation}
for one-forms $\omega dt$ commute with $H$.  We consider only those
$\omega(t)$ for which \ref{form} converges when acting on the part
of ${\cal H}$ corresponding to some spectral interval $[a,b]$ of $H$
that contains the eigenvalue zero.
Note that \ref{form}
does not depend on the choice of
lapse function $n(t)$.

Since $\omega dt$ is a
one-form, we assume that it satisfies the equation
\begin{equation}
-i {{\partial} \over {\partial t}} \Bigl( {{\omega (t) } \over
{N(t)}} \Bigr) = N(t) [H(t), {{\omega(t)} \over {N(t)}}].
\end{equation}
This is the case when $\omega$ is a sum of terms of the form:
\begin{equation}
N(t)
\prod_{j=1}^l ( {1 \over {N(t)}}
{{\partial} \over {\partial t}} )^{n_j}
\omega^{(j)}(t)
\end{equation}
and the $\omega^{(i)}(t)$ are built from the $Q(t)$'s and the
$P(t)$'s so that they satisfy
\begin{equation}
-i {{\partial} \over {\partial t}} \omega_j^{(i)}(t)
= N(t) [H(t), \omega^{(i)}(t)].
\end{equation}

To show that $\Omega$ commutes with the constraint, we consider
a basis of $H$-eigenstates $|E, k\rangle$ with
eigenvalue $E$ where $k$ is a discrete label that removes any
remaining degeneracy.  In general, these may not represent normalizable
states, but only states that are delta-function normalizable.
We take $\langle E_1, k | E_2, k_2 \rangle = \delta^{(?)}(E_1 - E_2)
\delta_{k_1, k_2}$ where this $\delta^{(?)}(E_1 - E_2)$
is to be interpreted as a delta-function or a Kronecker delta as appropriate.

Consider the matrix elements of ${{\omega(t)} \over {N(t)}}$
between any two
states of this type.  Note that
\begin{equation}
\label{om el}
\langle E_1, k_1 |{{\omega(t)} \over {N(t)}} | E_2, k_2 \rangle
= \exp\Bigl( i (E_1 - E_2) \int_{t'}^t  dt \ n(t) \Bigr)
\langle E_1, k_1 |
{{\omega(t')} \over {N(t')}} | E_2, k_2 \rangle
\end{equation}
so that we have
\begin{equation}
\label{sub pres}
\langle E_1, k_1 | \Omega | E_2, k_2 \rangle
=
2 \pi  \delta(E_1 - E_2) \langle E_1, k_1 | {{\omega (t')} \over {N(t')}}
| E_2, k_2 \rangle,
\end{equation}
using \ref{n cond 1}.
The delta-function in $E_1 - E_2$ guarantees that our operator
preserves the spectral subspaces of $H$, and therefore that
\begin{itemize}
\item{}
$[\Omega,H]=0$ when \ref{form} converges weakly.
\end{itemize}

The reader may be concerned by the presence of the delta-function in
$E_1 - E_2$ for the case where the spectrum of $H$ is discrete.
This concern appears to be valid and will be discussed in section
\ref{diss}.  Here, however, we will only be interested in applying
these techniques to the case where the spectrum of $H$ is purely
continuous (at least near the zero eigenvalue).

We now return to observables of the form \ref{qi}.  Our first task will
be to address the factor ordering of the integrand as
${{\partial} \over {\partial t}}
Q(t)$ will typically involve $P(t)$ and therefore fail to commute with
$Q(t)$ and $\delta(Q(t) - \tau)$.  From \ref{Omega action} and
\ref{it works}, we see that
any symmetric ordering of the factors
${{\partial} \over {\partial t}} Q(t)$, $Q(t)$, and $A(t)$ will
produce a Hermitian gauge invariant that we may take to be a
``quantization" of $[a]_{q = \tau}$.  However, experience shows that
the ordering
\begin{equation}
\label{ob def}
[A]_{Q = \tau} = {1 \over 2} \int_{-\infty}^{\infty} dt
\{ A(t), {{\partial} \over {\partial t}} \theta(Q(t) - \tau) \}_+
\end{equation}
is particularly convenient and leads to tractable calculations
and analyses.
As opposed to other factorizations that might be chosen \cite{Diss},
it has the
reassuring property (see \ref{RFP}) that $[{\rm sign}(P_0)]_{X^0
= \tau}$ is the
identity operator
for the free relativistic particle\footnote{Note that this is the
correct result as it is this operator that counts ``the number of
times that $x^0 = \tau$." Thus, $[{\rm sign}(p_0)]_{x^0=\tau}= 1$
in the classical case and $[1]_{x^0=\tau} = {\rm sign} (p_0)$.}.
Here, $\{X, Y \}_+ = XY + YX$ is the
anticommutator and $\theta(Q(t) - \tau)$ is the projection onto the
positive part of the spectrum of $Q(t) - \tau$.
We thus take \ref{ob def} as the definition of $[A]_{Q = \tau}$
in what follows.

Before deriving further
results, let us reflect for a moment on the two definitions of
``observable" mentioned above.  Within our scheme, the
canonical
idea that an ``observable" is an object that commutes with the
constraint (the Hamiltonian, $H$) does not capture the entire notion of
``gauge invariant" (see related comments in \cite{hisb,gp}).
For example, the object $N(t)$ commutes with the
constraint but does not
correspond to a gauge-invariant quantity.  Also, note that
the commutator of an operator ${\cal O}$ with $H$ is not
in general gauge invariant so that this definition is itself ``gauge
dependent."  For example, an argument similar to
that of \ref{om el} and \ref{sub pres}
shows that any object of the form
$\int_{-\infty}^{\infty} dt A(t)$, where $A(t)$ is an algebraic combination of
$Q_i(t)$ and $P_j(t)$, commutes with the Hamiltonian in a representation
for which $n(t) = 1$.  However, for more general choices of
lapse, it does not commute with $H$.  (This problem can be avoided
by requiring that $[{\cal O}, H] = 0$ in all irreducible representations
and not just be zero in the representation used.)
However, we have seen that reparametrization invariant
operators ${\it do}$ commute with the Hamiltonian.
Thus, we take reparametrization invariance as the more fundamental
notion of observable.

Finally, we give a few more definitions and useful results.
In order to study the case where $A(t)$ does
not commute with both $Q(t)$ and ${{\partial} \over {\partial t}} Q(t)$,
we introduce the operators:
\begin{equation}
\label{left}
[A]^L_{Q = \tau} = \int_{-\infty}^{\infty} dt
A(t) \Bigl( {{\partial} \over {\partial t}} \theta(Q(t) - \tau) \Bigr)
\end{equation}
and
\begin{equation}
\label{right}
[A]^R_{Q = \tau} = \int_{-\infty}^{\infty} dt
\Bigl( {{\partial} \over {\partial t}} \theta(Q(t) - \tau) \Bigr) A(t)
\end{equation}
which satisfy
\begin{eqnarray}
\label{g i def}
([A]^L_{Q = \tau})^{\dagger} = [A]^R_{Q = \tau}, \cr
[A]_{Q = \tau} = {1 \over 2} ([A]^L_{Q= \tau} + [A]^R_{Q = \tau}),
\end{eqnarray}
and, if $[A(t), {{\partial} \over {\partial t}} \theta(Q(t) - \tau)] = 0$,
we have
$[A]_{Q = \tau} = [A]^L_{Q = \tau} = [A]^R_{Q= \tau}$.

The convenient expressions
\begin{equation}
\label{calcL}
[A]^L_{Q = \tau} |E_,k;t'\rangle = 2 \pi i
\sum_{k^*} |E,k^*;t' \rangle \langle E,k^*;t'|A(t')
[H,\theta(Q(t') - \tau)] | E,k;t' \rangle
\end{equation}
and
\begin{equation}
\label{calcR}
[A]^R_{Q = \tau} |E_,k;t'\rangle = 2 \pi i
\sum_{k^*} |E,k^*;t' \rangle \langle E,k^*;t'|
[H,\theta(Q(t') - \tau)] A(t') | E,k;t' \rangle
\end{equation}
for the actions of these operators follow from \ref{sub pres} since
${1 \over N} {{\partial} \over {\partial t}} \theta(Q(t) - \tau)
= i [H, \theta(Q(t) - \tau)]$.  Here,
we have again assumed that $E$ lies in
the continuous spectrum of $H$.
In \ref{RFP} and \ref{sep} these relations will be used
to study $[A]_{Q = \tau}$.

\subsection{The Hilbert Space}
\label{in prod}

Equation \ref{sub pres}
is also sufficient for us to determine the physical inner product.
In the spirit of \cite{AA},
we wish to choose this inner product such that all $\Omega$ of the form
\ref{form} are Hermitian (or at least symmetric) when $\omega(t)$ is
hermitian.  Thus, we require
\begin{equation}
\label{req}
\overline{(|E= 0; k_1 \rangle , \Omega |E= 0, k_2 \rangle )_{phys}}
= (|E= 0; k_2 \rangle , \Omega |E= 0, k_1 \rangle )_{phys}
\end{equation}
where the overline denotes complex conjugation and $(.)_{phys}$ is the
physical inner product.
But, assuming that the spectrum of $H$ is continuous, we may
rewrite \ref{sub pres} as
\begin{equation}
\label{Omega action}
\Omega |E_1, k_1 \rangle = \sum_{k_2} |E_1, k_2 \rangle
\langle E_1, k_2 | {{\omega(t')} \over {N(t')}} | E_1, k_1 \rangle
\end{equation}
and we see that \ref{req} is satisfied if
\begin{equation}
\label{pip}
(|E= 0; k_2 \rangle , |E= 0, k_1 \rangle )_{phys}
= \delta_{k_2, k_1}
\end{equation}
since
\begin{equation}
\label{it works}
\overline{ \langle E=0, k_1 | {{\omega (t')} \over {N(t')}}
| E=0, k_2 \rangle  } = \langle E=0, k_2   | {{\omega (t')} \over {N(t')}}
| E=0, k_1 \rangle.
\end{equation}
We conclude that
\begin{itemize}
\item{}
All (convergent) observables of the form \ref{form} are symmetric on ${\cal
H}_{phys}$ with the inner product \ref{pip}\footnote{This
inner product has been independently introduced several times in
related contexts.  See \cite{ct,ah} for the cases known to the author.}.
\end{itemize}

Uniqueness of this inner product can be guaranteed by enforcing
\ref{req} for sufficiently many $\Omega$, such as those
($\Omega^{j_1 j_2}_{\pm}$)
defined
by any set of one-forms for which
\begin{equation}
\langle E= 0, k_2 | {{\omega^{j_1 j_2}_{\pm}(t')} \over {N(t')}} | E = 0, k_1
\rangle = (\delta_{k_1, j_1} \delta_{k_2, j_2} \pm \delta_{k_1, j_2}
\delta_{k_2, j_2}) i^{(1 \mp 1)/2}
\end{equation}
However, finding a set of physically interesting operators whose
Hermiticity makes \ref{pip} unique is a more difficult goal which we will
not pursue here.

\section{Simple Examples}
\label{se}

We now present two simple examples to illustrate how the formalism of
\ref{Dirac} may be applied and to show that it
produces reasonable results.  These test cases show that the methods
of \ref{Dirac} give the usual results for ``already deparametrized
systems" (\ref{ADS}) and lead to a familiar Hilbert space for the relativistic
free particle (\ref{RFP}).
These results validate the methods so that we may then
apply them to cosmological models in \ref{sep}.

In the examples we will see that our approach in fact provides an
``overcomplete set" of quantum observables.  By this we mean a set of
observables whose classical counterparts are overcomplete (see. for
example \cite{AA}) on the (constrained) phase space.

\subsection{Already Deparametrized Systems}
\label{ADS}

A standard testing ground for ideas regarding quantization of time
reparametrization invariant systems is the class of
systems that have been deparametrized (i.e., for which a time
function has been identified \cite{Kuchar}) or, what is nearly
equivalent, systems that (in the language of canonical
quantization) have a Hamiltonian constraint of the form
\begin{equation}
\label{constr}
p_0 + h_1 = 0
\end{equation}
where $p_0$ is some momentum and $h_1$ is independent of $p_0$.  For
such systems, the coordinate $q^0$ conjugate to $p_0$ effectively acts as
a clock and the system may be regarded as ``already deparametrized"
by the intrinsic time $q^0$.
A typical example is a parametrized nonrelativistic
particle \cite{KK} in which case \ref{constr} describes
such a system with Hamiltonian $h_1$ and
Newtonian time function $q^0$.  Note that $h_1$ may involve $q^0$, in which
case the corresponding Newtonian Hamiltonian is time-dependent.
We will see that the approach of \ref{Dirac} gives the usual
quantization
of nonrelativistic particles when applied to systems of this form.

We wish to
evaluate matrix elements of $[A]_{Q^0  = \tau}$ and begin by noting that
$Q^0(t)  =  Q^0(t') + T(t)$ where
$T(t) = \int_{t'}^t N(t)$ for some fixed $t'$.  Thus, we find
\begin{equation}
\label{IIIBaction}
\langle q^0_1, k_1;t' | [A]_{Q^0 = \tau} | q^0_2, k_2;t' \rangle
 =  \langle q^0_1, k_1;t'| A( t_{\tau - q^0_1})| q^0_2, k_2;t' \rangle
\end{equation}
where $t_{\tau - q^0_1}$ is the time $t$ when $T(t) = \tau-q^0_1$.
Equation \ref{IIIBaction} can be used directly
to show that $[A]_{Q^0=\tau}$ preserves physical
states, but this already follows from the general arguments of
\ref{Dirac}.

 From \ref{IIIBaction}, we see that
$[A]_{Q^0 = \tau} |\psi \rangle$ has components
\begin{equation}
\label{tau comps}
\langle q^0_1 = \tau - T(t), k_1; t | [A]_{Q^0 = \tau} |\psi \rangle
= \langle q^0_1 = \tau - T(t), k_1; t|A(t) | \psi \rangle
\end{equation}
on the $Q^0(t) = \tau - T(t)$ subspace.  Now, if
$|\psi \rangle$ is an eigenstate of $H$ with eigenvalue $E$, so is
$|\psi' \rangle = [A]_{Q^0 = \tau} |\psi\rangle$ and  it
satisfies
\begin{equation}
\label{diff}
-i {{\partial} \over {\partial q^0_1}} \langle q^0_1, k_1; t|\psi'\rangle
+ \langle q^0_1, k_1; t|H_1(t)|\psi'\rangle = E \langle q^0_1, k_1; t|\psi'
\rangle
\end{equation}
It is therefore completely determined by the components given in
\ref{tau comps}, much as is the case in, for example, \cite{atu}.
Since the action of $[A]_{Q^0 = \tau}$ on $|\psi \rangle$ is determined
through \ref{tau comps} and \ref{diff}, it follows that, for any
$Z_1(t),Z_2(t)$ built from the $Q(t)$'s and $P(t)$'s, if
$[Z_1(t),Z_2(t)] = Z_3(t)$ then we have
\begin{equation}
\label{IIIB inv comms}
\Bigl[ [Z_1]_{Q^0 = \tau}, [Z_2]_{Q^0 = \tau} \Bigr] = [Z_3]_{Q^0 = \tau}.
\end{equation}

We can also derive
\begin{equation}
\label{IIIB inv dyn}
-i {{\partial} \over {\partial \tau}} [Z_1]_{Q^0 = \tau}
= \Bigl[ \tilde{H}_1(\tau), [Z_1]_{Q^0 = \tau} \Bigr]
\end{equation}
by considering the matrix elements:
\begin{eqnarray}
\langle q^0_1 &=& \tau - T(t), k_1;t | -i{{\partial} \over {\partial \tau}}
[A]_{Q^0 = \tau} | \psi \rangle \cr
&=& -i{{\partial} \over {\partial \tau}}
\Bigl( \langle q^0_1 = \tau - T(t), k_1;t |
[A]_{Q^0 = \tau} | \psi \rangle  \Bigr)
 - \langle q^0_1 = \tau - T(t), k_1;t |P_0(t) [A]_{Q^0 = \tau} | \psi \rangle
\end{eqnarray}
and using $[H_1(t), Q^0(t) = 0$, $[P_0(t),A(t)] = 0$, $P_0(t) = H(t) -
H_1(t)$, and $[H(t),[A]_{Q^0=\tau}]=0$.
In \ref{IIIB inv dyn}, $\tilde{H}_1(\tau)$ is the operator built from
the $[Q]_{Q^0 = \tau}$'s and the $[P]_{Q^0=\tau}$'s in the same
way that $H_1$ is built from the $Q(t)$'s and the $P(t)$'s.
Thus, we have reproduced the usual results at the
level of algebras.  Since the states $|E,k_1\rangle$ which
satisfy $H|E,k_1\rangle = E$ and $\langle q^0,k_1|E,k_2\rangle = e^{iq^0E}
\delta_{k_1,k_2}$ form a basis for ${\cal H}$, this also follows
at the level of
the physical Hilbert space from our use of \ref{pip}.

\subsection{The Relativistic Free Particle}
\label{RFP}

Though hardly a sufficient test of a quantization scheme, every
procedure for
quantizing reparametrization invariant systems should be applicable
to the relativistic free particle.  We therefore take this as
our next special case and discuss the convergence
of the integrals \ref{left} and \ref{right}
and the operators they define.  We will also see
that the inner product \ref{pip} is unique, provided that a
simple set of observables are required to be symmetric.

We consider here a relativistic free particle in a flat Minkowski spacetime
of an arbitrary number of dimensions.
Because the equations of motion can be solved easily,
we can explicitly write the desired algebra\footnote{Note
that a ``deparameterized algebra'' based on
gauge fixing $x^0 = t$ may also be introduced for this
system and can be used in place of \ref{Xcom} to
construct the operators \ref{qi}.  A brief comparison of these
two possibilities, as well as some of the calculations
below, appeared in \cite{Diss}. }
as:
\begin{equation}
\label{Xcom}
[X^{\mu}(t), X^{\nu}(t')] = {i \over m}
\eta^{\mu \nu} \int_t^{t'} dt N(t), \qquad [N(t), X^{\mu}(t')] = 0
\end{equation}
where $\eta^{\mu \nu }$ is the Minkowski metric.  This follows
from \ref{dyn} in the form
\begin{equation}
\label{free dyn}
{d \over {dt}} P_{\mu}(t) = 0, \ {d \over {dt}} X_{\mu} = {{N(t)} \over
m} P_{\mu}
\end{equation}
so that $P_{\nu}(t) = P_{\nu}$.  The constraint is given by
$H = P^2 + m^2 = 0$.
As in \ref{Dirac}, $N(t)$ is some function $n(t)$
times the identity operator.

For each $t$, it is useful to introduce two bases
$\{|x;t\rangle \}$ and $\{|p;t\rangle\}$  that satisfy
\begin{eqnarray}
& X^{\mu}(t) |x;t \rangle = x^{\mu} |x;t \rangle , \
& P_{\nu} |p;t \rangle = p_{\nu} |p;t \rangle \cr
& \langle x; t| x' ; t \rangle = \delta(x - x'), \
& \langle p; t| p' ; t' \rangle = \delta(p - p') \exp(i \int_t^{t'} dt N(t)
 p_{\mu} p^{\mu}) \cr
& \langle x; t| p; t \rangle = {1 \over {\sqrt{2 \pi}}}e^{ipx}.
\end{eqnarray}

We then proceed to consider the operators $[A]_{X^0 = \tau}$.  Inserting
complete sets of states and
using $\int dE \sum_a |E,a \rangle \langle E,a| = \int dp |p;t'\rangle
\langle p,t'|$, $\delta(E - E^*) = \delta(p^2 - (p^*)^2)$, and
\begin{equation}
\langle p',t| \theta(X^0(t) - \tau) | p,t \rangle
= \delta(\overline{p} - \overline{p'}) \int_{\tau}^{\infty} {{dx^0}
\over {2 \pi}}
e^{i x^0 (p_0 - p'{}_0)}
\end{equation}
together with \ref{calcL}, we find that
\begin{eqnarray}
\label{sing}
[A]^L_{X^0 = \tau} | p;t'\rangle &=& -2 \pi i \int dp'{}_0 dp^*
\delta(p^2 - (p^*)^2) |p^*;t' \rangle (p_0^2 - p'{}_0^2)
\langle p^*;t'|A(t')|p'{}_0,\overline{p};t' \rangle
\cr &\times&
{ 1 \over 2} e^{i \tau (p'{}_0 - p_0)} [ \delta (p'{}_0 - p_0)
- {i \over { \pi (p'{}_0 - p_0})}].
\end{eqnarray}
Here, ${\overline p}$ is the collection of spatial components of $p$
so that $p = (p_0, {\overline p})$ and
the integration over $x^0$ has
been performed in a distributional sense.   Note, again, the presence of the
energy conserving delta-function which guarantees that our
operators commute with the constraint.

The expression \ref{sing} contains several
singular expressions, all of which are regulated by the factor
$(p'{}_0^2 - p_0^2)$ so that we find:
\begin{equation}
\label{reg}
[A]^L_{X^0 = \tau} |p;t'\rangle =
\int dp^* dp'_0  |p^*;t' \rangle
\delta (p^2 -(p^*)^2))
\langle p^* ; t|
A(t) |p'{}_0, \overline{p}; t \rangle
e^{i \tau (p'{}_0 - p_0)} [ p'{}_0 + p_0].
\end{equation}
If $A$ commutes with $P_0$, this simplifies to
\begin{eqnarray}
\label{comm case}
[A]^L_{X^0 = \tau} |p;t'\rangle &=&  \int d{\overline {p^*}}
\sum_{\epsilon \in \{+,-\}} |\epsilon \tilde{p}_0,{\overline {p^*}};t' \rangle
\langle {\overline {p^*}};t' |\overline{A(t')} |
\overline{p};t' \rangle_{\epsilon \tilde{p}_0}
e^{i \tau (\epsilon \tilde{p}_0 - p_0)}
{{[\epsilon \tilde{p}_0 + p_0 ] } \over { 2 \tilde{p}_0}}
\end{eqnarray}
where we have introduced the reduced matrix elements
$\langle \overline{p}';t'|{\overline{A(t')}}|
\overline{p}; t'\rangle_{p}$ that satisfy:
\begin{equation}
\label{rme}
\langle {p}';t'|A(t')| {p}; t'\rangle = \delta(p'{}_0 - p_0)
\langle \overline{p}';t'|{\overline{A(t')}}| \overline{p}; t'\rangle_{p_0}
\end{equation}
and where $\tilde{p}_0 = \sqrt{p^2_0 - \overline{p}^2 + \overline{p^*}^2}$.
If $A$ commutes with $Q^0$,  the reduced matrix elements \ref{rme}
do not depend on $p_0$.

Note that the
factor of $\epsilon \tilde{p}_0 + p_0$ in
\ref{comm case} minimizes the amount by which
this operator mixes positive and negative frequency states, though
some mixing still occurs.  Thus, the case of $A = X_i$
does not correspond to a Newton-Wigner operator (which would leave the
positive and negative frequency subspaces invariant).  On the other
hand, it may be checked that $[\theta(P_0) {\rm sign{P_0}} X^i]_{X^0 =
\tau}$ {\it is} just the Newton-Wigner operator $X^i_{NW}(\tau)$
associated with the coordinate $X^i$.

It now follows that Hermiticity of a few interesting operators
guarantees that the inner product \ref{pip} is unique.
{}From \ref{comm case}, we see that
\begin{equation}
[P_0\ {\rm sign}(P_0)]^L_{X^0 = \tau} |p;t'\rangle = p_0 |p;t'\rangle
\end{equation}
which is hermitian on ${\cal H}$ so that $[P_0\ {\rm sign}(P_0)]^L_{X^0 =\tau}
= [P_0\ {\rm sign}(P_0)]^R_{X^0 = \tau} = [P_0\ {\rm sign}(P_0)]_{X^0 =
\tau}$\footnote{A
similar calculation verifies that $[{\rm sign}(P_0)]_{X^0 = \tau} = \openone$
as claimed.}.
If we require $[P_0\ {\rm sign}(P_0)]_{X^0 = \tau}$
to be Hermitian on the physical space (spanned by $|
\epsilon |{\overline {p}}|, {\overline{p}};t' \rangle$),
it follows that the positive and negative frequency subspaces are
orthogonal.  If we also ask that $[X_i]_{X^0 = \tau}$ and
$[P_j]_{X^0 = \tau}$ ($i,j \in \{1,2,3\}$) be symmetric, then
the inner product is fixed up to an overall normalization on
the positive frequency
subspace and a similar normalization on the negative frequency subspace.
Additionally, if we require $[\Theta]_{X^0 = \tau}$ to be
Hermitian, where $\Theta$ is the time reversal operator ($\Theta|p;t\rangle
= |-p;t\rangle$), these normalizations must be identical and the inner
product is uniquely determined (up to an overal scale factor) to be:
\begin{equation}
\label{rfpip}
(|\epsilon |{\overline{p}}|, {\overline{p}} ; t' \rangle,
| \epsilon' |{\overline{p}}'|,  {\overline{p}}';t' \rangle)_{phys}
= {2 \sqrt{{\overline{p}}^2 + m^2}}{\delta({\overline{p}} -
{\overline{p}}') \delta_{\epsilon, \epsilon'}}
\end{equation}
Note that \ref{rfpip} agrees with \ref{pip} since $dE = 2|p_0| dp_0$.
It therefore follows from \ref{co} that all operators of the form
$[A]_{B = \tau}$ are symmetric (for Hermitian $A(t)$ and $B(t)$) and
we need not have chosen $x^0$ as the ``intrinsic clock."
Finally, we verify that we have constructed an overcomplete
set of observables as densely defined operators on ${\cal H}_{phys}$.
This follows since \ref{rfpip} shows
that $[X_i]_{X^0 = \tau}$, $[P_j]_{X^0 = \tau}$,
$[P_0\ {\rm sign}(P_0)]_{X^0 = \tau}$ and $[\Theta]_{X^0 = \tau}$
yield normalizable states when acting on
any $|\psi\rangle \in {\cal H}_{phys}$ for which
$(|\epsilon |{\overline{p}}|, {\overline{p}} ;
t' \rangle, |\psi\rangle)_{phys}$
is a sufficiently smooth and rapidly decreasing function of $\overline{p}$
and thus are densely defined.

\section{Recollapsing Dynamics: Separable Semi-bound Models}
\label{sep}

The LRS Bianchi IX and Kantowski-Sachs
minisuperspaces fall into the class of models
that we refer to as ``separable and semi-bound."  These are, roughly
speaking, models for which the potential is separable and which
(classically) describe spacetimes whose homogeneous slices expand,
reach a maximum volume, and recontract.  They are thus ``semi-bound" in the
sense that they do not classically reach arbitrarily large size.

Below, we apply the methods of section \ref{Dirac} to such models.  As before,
we verify the convergence (\ref{ic}) of the integrals that define our
observables and construct a complete set of densely defined operators
on ${\cal H}_{phys}$ (\ref{cs}).  In addition, we derive quantum analogues of
\ref{class recol} and \ref{all a}, showing that our quantization also predicts
recollapsing behavior.

\subsection{The System and Observables}

Separable semi-bound systems
are time-reparametrization invariant
models that can be described in the canonical formalism with
a Hamiltonian constraint of the form $h = h_0 - h_1 = 0$ where
$h_0$ is of the form $p_0^2 + V_0(q^0)$ for some canonically conjugate
pair $(p_0,q^0)$ which take values in $(-\infty,\infty)$,
$h_1$ does not involve either $p_0$ or $q^0$, and $h_1$
is a Hamiltonian of the type appropriate to a
nonrelativistic particle.  Thus, we
consider the case where the configuration space ${\cal Q}$ is of the
form ${\bf R} \times {\cal Q}_1$ and the phase space is
$T^*{\cal Q} = T^*{\bf R} \times T^*{\cal Q}_1$.  In addition,
we will ask that $V(q^0)$ be smooth, that $V_0(q^0) \rightarrow \infty$
as $q^0 \rightarrow + \infty$, $V_0(q^0) > 0$,
and $V_0(q^0) < W e^{\alpha q_0}$
for some positive $W$ and $\alpha$.  This last condition is important
only for large negative $q^0$.  This leads to the following:

\begin{itemize}
\item{}
{\em Definition:}  A separable semi-bound system is a Hilbert
space ${\cal H}_{aux} = L^2({\cal R}) \otimes {\cal H}_1$ together
with a self-adjoint operator $H = H_0 - H_1$ such that
$H_1 = \openone \otimes \tilde{H}_1$ for some $\tilde{H}_1$
and $H_0 = \tilde{H}_0 \otimes \openone$ for $\tilde{H}_0
= - {{\partial^2} \over {\partial (q^0)^2}} + V(q_0)$ and a
potential $V(q^0)$ which satisfies (on every semi-bounded
interval $(-\infty, \tilde{q}^0)$) the bound $V(q^0)
< W e^{-\alpha q^0}$ for some $W, \alpha >0$, but for which
$V(q^0) \rightarrow \infty$ as $q^0 \rightarrow \infty$.
\end{itemize}

These conditions imply that,
when viewed
as an operator on the Hilbert space $L^2({\bf R})$ in a ``$q^0$ - coordinate
representation,'' the spectrum of $H_0$ is purely continuous.
The condition that
$V_0(q^0) \rightarrow \infty$ will allow us to work with convergent
expressions that require no regularization or careful treatment.
 From \cite{atu}, we see that when described by the appropriate variables and
with a rescaled Hamiltonian, the LRS Bianchi IX and Kantowski-Sachs models
may be placed in this form with $q^0 = \ln( \det g)$ where $g$
is the 3-metric of a homogeneous slice.

As described in \ref{co},
we use the auxiliary Hilbert space $L^2({\cal Q})$ to carry a
representation of the algebra defined by \ref{quant alg} and \ref{dyn} and
require that $H = H_0 - H_1$ be a Hermitian factor ordering of $h$
on this space (and similarly for $H_0$ and $H_1$).  From this and
the above conditions, it follows that $H$ has a purely continuous
spectrum with $\sigma_c(H) = (-\infty, \infty)$.
Note that $H$ is time independent and that, since they
commute, $H_0$ and $H_1$ are time independent as well.

As before, we will make use of a basis in this
auxiliary space that is tailored to our system.
This basis will consist of (delta-function normalizable) states
$\{ | E_0, E_1, k \rangle \}$ that are eigenvectors of $H_0$ with
eigenvalue $E_0$ and that are eigenvectors of $H_1$ with eigenvalue
$E_1$.  The (discrete) label $k$
is to remove any remaining degeneracy in our
description.  Thus, $H |E_0, E_1,k \rangle = (E_0 - E_1) |E_0, E_1, k
\rangle$.
Note that this set of labels leads to a decomposition of
our auxiliary Hilbert space as a direct product ${\cal H}
= {\cal H}_0 \otimes {\cal H}_1$ through
$|E_0, E_1, k \rangle = |E_0 \rangle \otimes |E_1, k \rangle$ since,
for systems of this type, the spectral subspaces of $H_0$ on
$L^2(q^0)$ are nondegenerate.

Also as in \ref{co}, the (constrained) Hamiltonian $H$ and
the lapse $N(t)$ define a parameter time evolution that, for the
canonical variables, is given by
\begin{equation}
\label{Aev}
A(t) = \exp(i H \int_{t'}^t dt N(t) )
A(t') \exp(-i H \int_{t'}^t dt N(t) ).
\end{equation}
In irreducible
representations, $N(t)$ is proportional to the identity: $N(t) = n(t)
\openone$ and we assume that \ref{n cond 1}
holds.
Also, for each $t$,
our representation has a set of states $|q;t\rangle $ labeled by points
$q$ in the configuration space ${\cal Q}$  that define a ``configuration
representation at time $t$" and satisfy $\langle q;t|q';t\rangle
= \delta(q, q')$.  As in
the case of the relativistic free particle, it will be convenient to
introduce a family of ``energy bases" $\{ |E_0, E_1,k ; t \rangle \}$
(also labeled by $t$), where
\begin{equation}
|E_0, E_1,k  ; t \rangle = \Bigl( \exp [
i (E_0 - E_1) \int_{t'}^t dt N(t)
] \Bigr)  |E_0, E_1, k ; t' \rangle
\end{equation}
so that inner products of the form $\langle q ;  t | E_0, E_1,k
; t \rangle$
are independent of $t$.  The arbitrary phase in the definition of
$|E_0, E_1,k \rangle$ may be used to set $|E_0, E_1,k \rangle
= |E_0, E_1,k ; t' \rangle$ for any $t'$.

Since the spectrum of $H$ includes zero, the set of physical
states that are delta-function normalizable in the
auxiliary inner product and satisfy $H|\psi\rangle = 0$ is not empty.
They are of the form $|E_0 = E_1,E_1,k;t' \rangle$ where $E_1$ ranges
over the spectrum of $H_1$ and $k$ ranges over its full set of values.
 From \ref{pip}, the physical inner product is just
\begin{equation}
(|E^*_1,E^*_1,k^*;t'\rangle, |E_1,E_1,k;t'\rangle)_{phys}
= \delta^{(?)}(E_1^*,E_1) \delta_{k^*,k}
\end{equation}
where the $\delta^{(?)}(E^*_1,E_1)$ is to be interpreted as a delta-function
or Kronecker delta as appropriate to the spectrum of $H_1$.

A few brief comments are now in order before deriving the rigorous
convergence and recollapse results.
We again wish to consider the action of the operators $[A]_{B = \tau}$.
Let us suppose that $B$ is, in fact, a configuration variable.  Note that
\begin{equation}
\langle E'{}_0, E'{}_1, k'; t| \theta(B(t) - \tau) | E_0 , E_1, k ; t \rangle
= \int_{{\cal Q}; b(q) > \tau} dq \langle E'{}_0, E'{}_1,k'; t|q;t\rangle
\langle q;t  | E_0 , E_1,k ; t \rangle
\end{equation}
so that we may write
\begin{eqnarray}
\label{lelm}
\langle E^*_0, E^*_1, k^*; t&|&[A]^L_{B = \tau} | E_0 , E_1,k ; t \rangle
\cr &=& -i 2 \pi \delta(E_0 - E_1 - E^*_0 + E^*_1)
\int dE'{}_0 \sum_{E'{}_1, k'}
(E_0 - E_1 - E'{}_0 + E'{}_1) \cr &\times&
\langle E^*_0, E^*_1,k^*;t |A(t) | E'{}_0, E'{}_1,k';t \rangle
\int_{{\cal Q}; b(q) > \tau} \langle E'{}_0, E'{}_1,k'; t|q;t\rangle
\langle q;t  | E_0 , E_1,k ; t \rangle
\end{eqnarray}
and, similarly for $[A]^R_{B=\tau}$.

Two cases are now of interest.  The first occurs when the spectrum of
$H_1$ is entirely discrete.  Then, the factor states $|E_1 \rangle$
are normalizable and the wavefunctions
$\langle E'{}_0, E'{}_1,k'; t|q;t\rangle$
fall off rapidly in the directions in which ${\cal Q}$ is not compact
and $q^0$ remains constant.  Furthermore, because of the assumption that
$V_0(q^0) \rightarrow \infty$ as $q^0 \rightarrow \infty$, these
wavefunctions also fall off in the positive $q^0$ direction.  Thus, if
$b(q)$ is any configuration variable for which  $q^0$
is bounded from below in
the region of ${\cal Q}$ with $b(q) \ge \tau$, this factor represents
a convergent expression.
At the other extreme, if the spectrum of $H_1$ is purely continuous (as
in \ref{RFP}), then
this integral will converge only in a distributional sense, as is
appropriate.  Note that convergence of these integrals guarantees that
the matrix element in \ref{lelm} (and correspondingly for
$[A]^R_{Q^0=\tau}$ vanishes in the
large $\tau$ limit, verifying the recollapsing behavior of \ref{all a}
in a weak sense.

\subsection{Rigorous Convergence and Recollapse Results}
\label{ic}

We now present two results concerning the convergence
of the sums in expressions \ref{calcL} and \ref{calcR}.  These
results will show rigorously that a quantum analogue of
\ref{all a} holds.  As usual, the stronger
result follows for the simpler case.

Let us begin by
assuming that $B=Q^0$ and that
$A(t)$ commutes with both $Q^0(t)$ and $P_0(t)$ so that
$[A]^L_{Q^0 = \tau} = [A]^R_{Q^0 = \tau} = [A]_{Q^0 = \tau}$.
Thus, we may again introduce the reduced matrix elements
$\langle E'{}_1,k' | \tilde{A} | E_1,k \rangle$ that satisfy
\begin{equation}
\label{gen red}
\langle E'{}_0, E'{}_1,k';t' | A(t) | E_0, E_1,k ; t'\rangle =
\delta (E'{}_0 - E_0)
{}_1\langle E'{}_1,k' | \tilde{A} | E_1,k \rangle_1
\end{equation}
and are independent of $E_0$.
Note that \ref{gen red} defines a ``reduced operator" $\tilde{A}$ on the
factor space ${\cal H}_1$.
The action of $[A]_{Q^0 = \tau}$ now takes the form:
\begin{eqnarray}
\label{stc}
[A]_{Q^0 = \tau} |E_0, E_1,k;t'\rangle &=&
-i 2 \pi \sum_{E^*_1,k^*}
 (E^*_1 - E_1 )
|E_0 - E_1 + E_1^*, E^*_1,k^* ; t' \rangle_1 \
{}_1\langle E^*_1,k^* | \tilde{A} |E_1,k; \rangle_1
\cr &\times& \int_{q_0 > \tau} dq_0 \
{}_0\langle E_0 - E_1 + E^*_1 |q_0 \rangle_0 \
{}_0 \langle q_0 | E_0 \rangle_0
\end{eqnarray}
where
bras and kets with subscripts $0$ and $1$ refer to states in ${\cal H}_0$
and ${\cal H}_1$ respectively.
Note that while all terms in this sum are finite it
remains to show that the sum itself represents a normalizable state in the
physical space.  Note that this sum will be at best
delta-function normalizable
in the auxiliary space since $|E_0, E_1, k \rangle$ is only
delta-function normalizable.

In order to discuss the convergence of the sum in \ref{stc}, we
would like to find a bound $M(\tau, E)$ such that
$|I(E, E^*, \tau)| \leq  M(\tau, E)$ where
\begin{equation}
\label{Idef}
I(E,E^*, \tau) = \int_{q^0 \geq \tau}
\langle E_0 = E^* |q^0 \rangle \langle q^0 | E_0 = E\rangle.
\end{equation}
We use the fact that the states are delta-function normalized
so that
$I(E,E^*,\tau) = \delta(E - E^*) -  \int_{q^0 \leq \tau}
\langle E^* |q^0 \rangle \langle q^0 | E \rangle$.
Because the potential $V_0(q^0)$ is exponentially small for large
negative $q^0$, we approximate the function $\langle q^0|E \rangle$
by plane waves through the expression:
\begin{equation}
\label{Delta def}
\langle q^0|E \rangle =
{1 \over {\sqrt{4 \pi p} }}  \Bigl( e^{ipq^0} + \alpha_E e^{-ipq^0}
\Bigr) + {1 \over {\sqrt{p}} } \Delta_E(q)
\end{equation}
for the appropriate $\alpha_E$ with $|\alpha_E| = 1$
and some $\Delta_E$ which we
expect to be small.  Here, we have set $p = \sqrt{E}$.
Note that we have chosen to normalize the plane
waves in a manner consistent with the normalization $\langle E | E'
\rangle = \delta(E-E') = {1 \over{2p}} \delta(p - p')$.

Thus, we may expand $I$ as
\begin{eqnarray}
\label{I}
I = \delta(E - E^*) &-& {1 \over {2p}} \delta(p - p^*)
+ {i \over {4 \pi (p + p^*)}} {(
e^{i \tau ( p+p^*)}\alpha_E - e^{-i \tau(p+p^*)}
\overline{\alpha_{E^*}}) \over
{\sqrt{p p^*}}} \cr
&+& {i \over {4 \pi (p^* - p)}} {{e^{i \tau(p^* - p)}
- e^{-i \tau(p^* - p)} \alpha_E
\overline{\alpha_{E^*}}}
\over {\sqrt{p p^*}}} \cr &-& {1 \over {\sqrt{4 \pi p p^*}}}
\Bigl( \tilde{\Delta}^{\tau}_E(-p^*) +
\overline{\alpha_{E^*}} \tilde{\Delta}^{\tau}_E(p^*)
+ \overline{\tilde{\Delta}^{\tau}_{E^*}(-p)} + \alpha_E
\overline{\tilde{\Delta}^{\tau}_{E^*}(p)} \Bigr) \cr &-& \int_{q_0 \leq \tau}
dq_0 {{\Delta_E \overline{\Delta_{E^*}}} \over {\sqrt{ p p^*}}}
\end{eqnarray}
where the bars denote complex conjugation and
\begin{equation}
\tilde{\Delta}^{\tau}_E(p^*) = \int_{q^0 \leq \tau} e^{iq^0p^*}
\Delta_E(q^0)
\end{equation}
The first two terms in \ref{I} cancel.

The bound on the
potential may now be used to verify
that $\Delta_E$ can be regarded as small.
By the usual arguments, it follows that $\Delta_E$
satisfies
\begin{equation}
\Bigl( {{\partial^2} \over {\partial (q^0)^2}} + E ) \Bigr) {{\Delta_E}
\over {\sqrt{p}}}
= V_0(q^0)  \langle q_0| E \rangle
\end{equation}
which has the solution
\begin{equation}
\label{sol}
\Delta_E(q^0) = \sqrt{p} e^{ipq^0} \int_{-\infty}^{q^0} dq'{}^0 e^{-2ipq'{}^0}
\int_{-\infty}^{q'{}^0} dq''{}^0 e^{ipq''{}^0} \langle q''{}^0|E \rangle
V_0(q''{}^0)
\end{equation}
since, for the right choice of $\alpha_E$,
$\Delta_E$ satisfies the boundary conditions $\Delta_E
\rightarrow 0$ and ${{\partial} \over {\partial q^0}} \Delta_E
\rightarrow 0$ as $q^0 \rightarrow - \infty$.  Let us rewrite this
expression using \ref{Delta def}:
\begin{eqnarray}
\label{approx}
\Delta_E(q^0) &=& {{e^{ipq^0}} \over {\sqrt{4 \pi}}}
\int_{-\infty}^{q^0} dq'{}^0 e^{-2ipq'{}^0}
\int_{-\infty}^{q'{}^0} dq''{}^0
e^{2ipq''{}^0} V_0(q''{}^0)\cr
&+&
{{e^{ipq^0}} \over {\sqrt{4 \pi}}}
\int_{-\infty}^{q^0} dq'{}^0 e^{-2ipq'{}^0}
\int_{-\infty}^{q'{}^0} dq''{}^0
\alpha_E V_0(q''{}^0)\cr
&+&  e^{ipq^0} \int_{-\infty}^{q^0} dq'{}^0 e^{-2ipq'{}^0}
\int_{-\infty}^{q'{}^0} dq''{}^0 e^{ipq''{}^0} \Delta_(q''{}^0) V_0(q''{}^0)
\end{eqnarray}

Let $\beta_1(E)$ and $\beta_2(E)$ be the maximum values of the first and
second terms in \ref{approx} for $q^0 \in (-\infty, \tau]$, which exist since
$V(q^0)$ is smooth and
bounded on this interval by $W
e^{\alpha q^0}$.  Note that $\beta_1(E), \beta_2(E)
\rightarrow 0$ as $E \rightarrow \infty$ since
both describe integrals of fixed $L^p(-\infty, \tau)$
functions against oscillating exponentials.  Here, the integral over $q'{}^0$
is highly oscillatory in the definition of
$\beta_2(E)$ and the integral over $q''{}^0$
is relevant for $\beta_1(E)$.

Further, since $\langle
q|E \rangle$ is smooth and bounded as $q \rightarrow - \infty$,
$|\Delta_E|$ takes some maximal value $\Delta_E^M$
on $(-\infty,\tau)$ by \ref{sol} and we see that
$\Delta_E^M \leq \beta_1(E) + \beta_2(E) + \gamma \Delta_E^M$,
where $\gamma = {W \over {\alpha^2}} e^{\alpha \tau}$
is independent of $E$.  Let us
suppose for the moment that $\gamma < 1$ so that
\begin{equation}
\label{beta bound}
\Delta^M_E \leq {{\beta_1(E) + \beta_2(E)} \over {1 - \gamma}}.
\end{equation}
Thus,
\begin{equation}
\label{est}
|\Delta_E(q^0) -
{{e^{ipq^0}} \over {\sqrt{4 \pi}}}
\int_{-\infty}^{q^0} dq'{}^0 e^{-2ipq'{}^0}
\int_{-\infty}^{q'{}^0} dq''{}^0 (e^{2ipq''{}^0}
+ \alpha_E)
V_0(q''{}^0)| \leq
{{(\beta_1(E) + \beta_2(E)) \gamma} \over {1 - \gamma}}.
\end{equation}
Since $\beta_1 + \beta_2 \rightarrow 0$,
this estimate
is accurate in the $E \rightarrow \infty$ limit
and $\Delta_E(q^0) \rightarrow 0$.  A similar
argument using \ref{approx} and \ref{beta bound} demonstrates that
\begin{equation}
\label{D lim}
\int_{-\infty}^{\tau} dq^0 {{|\Delta_{E^*}(q^0)|} \over {\sqrt {p^*}}}
\rightarrow 0
\end{equation}
in the $E^* \rightarrow \infty$ limit and that the $L^2$ norm of
${{\Delta_{E^*}} \over {\sqrt{p^*}}}$ on
$(-\infty, \tau)$ is finite and also vanishes for large $E^*$.
It follows that $I(E,E^*,\tau) \rightarrow 0$
as $E^* \rightarrow 0$ and thus that there exists a bound
$M(E,\tau) \geq |I(E,E^*,\tau)|$.  To see this, recall that we know that
$I(E,E,\tau)$ is finite so that we may ignore the apparent divergences
in \ref{I} for $p^* = \pm p$.  Now, for $p\neq p^*$ $I$ is a continuous
function of $E^*$ by \ref{I} and \ref{sol}.
The first two terms in \ref{I} cancel and the next two
terms are explicitly decreasing for large $p^*$.  The pair of
terms involving $\overline{\tilde{\Delta}_{E^*}^{\tau}(\pm p)}$ are
bounded by a multiple of \ref{D lim} and the terms
$\tilde{\Delta}_{E^*}^{\tau}(\pm p^*)$ vanish as they are the integral
of a smooth $L^2$ function against a rapidly oscillating exponential.
The last term is bounded by the product of ${{\Delta^M_E}
\over {\sqrt{p }}}$ and \ref{D lim}.
Thus, we have derived the existence of a bound (and in fact
of a maximal value) for $|I(E,E^*,\tau)|$, assuming $\gamma < 1$.

We now consider the more general case and show that, by repeating
the approximation described by \ref{approx}, we can follow a
similar proof without assuming $\gamma < 1$.  Note that if
we iterate this approximation $n$ times, replacing at each step the
$\Delta_E$ that appears in the integral on the right hand side of
\ref{approx} with the entire right hand side, then the term on the
right hand side that still contains $\Delta_E$ can be bounded
by an expression of the form:
\begin{equation}
\label{new bound}
\Delta_E^M
\int_{\tau \leq q_1 \leq q_2 \leq q_3 \leq . .  \leq q_{2n}}
d^{2n}q \prod_{j = 1}^n V(q_{2j}) \leq {{W^n} \over {(n!)^2 \alpha^{2n}}}
e^{n \alpha \tau}  \Delta^E_M
\end{equation}
and so approaches zero as $n \rightarrow \infty$.  Hence, for any $V_0$
of the form specified, there is some finite $n$ such that the bound in
\ref{new bound} (which we will call $\gamma_n\Delta^E_M$) is less than
$\Delta^E_M$  so that $\gamma_n < 1$.
A bound of the form \ref{beta bound} again follows with $\gamma$ replaced
by $\gamma_n$ and $\beta_1$, $\beta_2$
replaced by the appropriate $\beta_{1,n}$
and $\beta_{2,n}$
which also vanish in the $E \rightarrow \infty$ limit.  The
rest of the argument proceeds as before, and we conclude that
$I(E,E^*,\tau)$ is a bounded function of $E^*$; that is,

\begin{itemize}
\item{}
There is a function
$M(E,\tau)$ such that $M(E,\tau) \geq |I(E,E^*, \tau)|$.
\end{itemize}

{}From \ref{stc}
and our bound $M(E,\tau)$, we now see that
$[A]_{Q = \tau} |E_0, E_1, k ;t\rangle$ is a normalizable state for any
$A$ such that
\begin{equation}
\label{norm cond}
\sum_{E^*_1,k^*} (E_1^* - E_1)^2 |\langle E^*_1,k^*
|\tilde{A}|E_1,k \rangle |^2
\leq \infty$, i.e., such that
$||[\tilde{A},\tilde{H}_1]|E_1,k\rangle||_{{\cal
H}_1} < \infty.
\end{equation}
where $||(|\psi\rangle)||_{{\cal H}_1}$ represents the norm of the state
$|\psi \rangle$ in the factor space ${\cal H}_1$.  This allows us the
conclude the following result:

\begin{itemize}
\item{}
{\em Result 1:} Let $A$ be a densely defined operator of the form
$\openone \otimes \tilde{A}$ on ${\cal H}_{aux}$ of a separable
semi-bound system.  Further, let $A(t)$ be the one parameter
family of operators defined through \ref{Aev} with $A(0) \equiv A$, and
let $[A]_{Q^0 = \tau}$ be the one parameter family of operators on
${\cal H}_{phys}$ defined by \ref{g i def}, \ref{calcL}, and \ref{calcR}.
Note that our construction (\ref{pip}) of ${\cal H}_{phys}$
identifies ${\cal H}_{phys}$ with ${\cal H}_1$ at $t = 0$.
Then, under this identification, if $|\psi\rangle \in {\cal H}_1$
lies in the domain of $[\tilde{A},H_1]$, it also lies in the
domain of $[A]_{Q^0 = \tau}$.  In particular, if $[\tilde{A},H_1]$
is densely defined, so is $[A]_{Q^0 = \tau}$.
\end{itemize}

It now follows that
quantization of ``separable semi-bound models" captures
the classical notion of ``recollapse" expressed through
\ref{all a} for $a(t)$ independent of $q^0(t)$ and
$p_0(t)$.
As in \cite{tril}, this is to be expected since we have used
the auxiliary space $L^2({\cal Q})$.
Below, we show that the norms of all states of the form $[A]_{Q^0 = \tau}
|E_0,E_1,k;t'\rangle$ vanish in the large $\tau$ limit and thus that
the operator $[A]_{Q^0 = \tau}$ converges strongly to
zero.

Note that we may take
the bound $M(E,\tau)$ on the integral $I$ to be given by the
maximum value of $I(E,E^*,\tau)$ over $E^*$, which exists by
our previous argument.  Thus, $M(E,\tau)$ is also given by
an expression of the form $|\int_{q^0 \geq \tau} f(E)|$ for $f \in L^1(\tau,
\infty)$.
Since this integral converges, it must vanish in the large
$\tau$ limit and, since the norm of $[A]_{Q^0 = \tau} |E_0, E_0,a;t' \rangle$
in ${\cal H}_{phys}$
is bounded by the product of $M^2$ and the norm given in \ref{norm cond},
we find

\begin{itemize}
\item{}
{\em Result 2:} For $A$ as in Result 1,
$[A]_{Q^0=\tau} \rightarrow 0$ as $\tau \rightarrow
\infty$ in the sense of convergence on a dense set of states; that is,
$[A]_{Q^0=\tau}$ converges strongly to the zero operator.
\end{itemize}

Unfortunately, considering only operators $A$ that commute with
$Q^0$ and $P_0$, is not sufficient to conclusively
demonstrate recollapsing behavior.
Recall that if a classical spacetime $s$ expands smoothly and then
recollapses, $[a]_{q^0 = \tau}(s)$ represents a difference of a term
associated with the expansion and a term associated with the
collapse.  Thus, our discussion to this point leaves
open the possibility that this quantization should be interpreted
as describing pairs of contracting and expanding spacetimes such that
the appropriate terms cancel for large $\tau$.
It would be comforting to find that operators insensitive
to the expansion or contraction of the universe vanish in the
large $\tau$ limit as well.

As suggested in section \ref{class}, we
would like to study the operator $[\rm{sign}(P_0)]_{Q = \tau}$.  However,
for technical reasons we will instead work with
\begin{equation}
\label{op def}
{\cal O}_{\tau} \equiv \Bigl[ {1 \over {H_0+1}}
\theta(Q^0 - \tau')\rm{sign}(P_0) \theta(Q^0 - \tau'){1 \over {H_0+1}}
\Bigr]_{Q^0 = \tau}
\end{equation}
and derive the analogue of \ref{all a} for this operator.
Note, first, that the factors of $\theta(Q - \tau')$ are
completely irrelevant classically for $\tau' < \tau$.  Despite this, they
will greatly simplify the details in what follows.
Also, $H_0$ is time independent and the classical $h_0$ is constant
along a solution.
Thus, \ref{op def} is nearly as satisfactory as
$[\rm{sign}(P_0)]_{Q = \tau}$ since,
if the corresponding classical observable approaches zero along a
solution for
large $\tau$, it tells us that the spacetime in question does not
reach arbitrarily large values of $q^0$.

Now, for an operator $A$ (such as \ref{op def})
that is constructed {\it only} from $Q^0$ and $P_0$,
the matrix element
$\langle E_0^*, E^*_1, k^* ;t'|A(t')|E_0, E_1, k ;t' \rangle$
must vanish unless $E^*_1 = E_1$ and $k^* = k$.  Thus,
$[A]_{Q^0 = \tau}|E_0,E_1,k;t'\rangle$ must be again
proportional to $|E_0, E_1, k;t'\rangle$.  From \ref{calcL} and
\ref{calcR}, we see (if the spectrum of $H_1$ is
discrete) that in fact
\begin{eqnarray}
\label{ev}
[A]_{Q^0 = \tau}|E_0,E_1,k;t'\rangle &=& 2 \pi
|E_0,E_1,k;t'\rangle \cr &\times&
{\rm{Im}}  \langle E_0, E_1, k;t'| A(E_0 - H_0)
\theta(Q^0(t') - \tau)|E_0, E_1, k;t' \rangle
\end{eqnarray}
where $\rm{Im}$ denotes the imaginary part.
If the spectrum of $H_1$ is continuous, the
action of $[A]_{Q = \tau}$ on $|E_0,E_1,k;t\rangle$ will not, of course,
produce a state that is normalizable in the physical inner
product.  However, essentially the same analysis follows by
considering wave packets and using the fact that $A(E_0-H_0)
\theta(Q-\tau)$ is of the form ${\cal O} \otimes \openone$ with
respect to the decomposition ${\cal H} = {\cal H}_0 \otimes {\cal H}_1$.

Thus, our task is to show that this expectation value is finite
and vanishes as $\tau \rightarrow \infty$.  Note that
$\theta(Q^0(t') - \tau)|E_0,E_1, k;t' \rangle$ is actually
normalizable in the auxiliary inner product since
$\langle q^0|E\rangle$ decays faster than
exponentially for large positive $q^0$.
We may express the expectation value in \ref{ev} for the
case where $A = {1 \over {H_0 +1}} \theta(Q -\tau') {\rm{sign}}(P_0)
\theta(Q - \tau') {1 \over {H_0+1 }}$ as
\begin{equation}
\label{finite}
{1 \over {E_0 + 1}} \langle \psi(\tau')| {\rm{sign}}(P_0(t'))
\theta(Q^0(t') - \tau){{E_0 - H_0} \over
{H_0 + 1}} |\psi(\tau)\rangle
\end{equation}
where $|\psi(\tau)\rangle = \theta(Q^0(t') - \tau) |E_0,E_1,k;t'\rangle$
and similarly for $|\psi(\tau')\rangle$.   Note that \ref{finite}
is the matrix element of a bounded operator between two normalizable
states and is therefore finite and bounded by some multiple
of the norm of $|\psi(\tau)\rangle$.  Again, $[A]_{Q = \tau}$
is densely defined on our physical Hilbert space.  Also, the norm
of $|\psi\rangle$ decreases to zero as $\tau \rightarrow \infty$,
so the action of $[A]_{Q = \tau}$ vanishes in this limit on any
finite combination of the states $|E_0,E_1,a;t'\rangle$.
Thus we find:

\begin{itemize}
\item{}
{\em Result 3:} ${\cal O}_{\tau}$ of \ref{op def} is
densely defined and converges
strongly to zero as $\tau \rightarrow \infty$.
\end{itemize}

The statement that there may be states to which this
does not apply (i.e., that are not in the appropriate dense set) is of
the same type as the statement that there are states in nonrelativistic
quantum mechanics with infinite expectation value of the energy.
For this reason, we feel justified in saying that
when $Q^0$ has the interpretation of a scale factor for a cosmology,
our quantization predicts that such spacetimes recollapse.
Note that this rules out the possibility of an ``evolution law" of the
type \ref{IIIB inv dyn}.  There is no exact
unitary evolution in $\tau$-time for
semi-bound separable systems, just as there is no corresponding
exact Hamiltonian evolution at the classical level.

\subsection{A Complete Set of Operators}
\label{cs}

Our final result will be that a complete set of densely defined
quantum observables can be constructed by the
techniques presented above.
In order to prove this general result, we
consider operators defined somewhat differently than in \ref{left}
and \ref{right}.
We will also assume that, for each $E_0$, $E_1$,
the label $k$ takes only a finite set of values.  This is the case
for both LRS Bianchi IX and the Kantowski-Sachs Model.

For any operator $B(t) = B(Q(t), P(t))$
bounded on ${\cal H}$,
consider the objects

\begin{eqnarray}
\label{bounded version}
\widetilde{[B]}^L_{Q^0 = \tau} &=& \int_{-\infty}^{\infty} dt
f(H_0) {1 \over {H+i}} \theta(Q^0(t) - \tau') B(t) \theta(Q^0-\tau')
{1 \over {H-i}} \Bigl( {{\partial} \over {\partial t}} \theta(Q^0 - \tau)
\Bigr) f(H_0)
\cr
\widetilde{[B]}^R_{Q^0 = \tau} &=& \int_{-\infty}^{\infty} dt
f(H_0) \Bigl( {{\partial} \over {\partial t}} \theta(Q^0 - \tau)\Bigr)
{1 \over {H+i}} \theta(Q^0(t) - \tau') B(t) \theta(Q^0-\tau')
{1 \over {H-i}}
f(H_0)
\end{eqnarray}
and note that these satisfy the analogues of \ref{g i def} when $B(t)$
is Hermitian on ${\cal H}$.
Here, $f$ is an as yet unspecified real function that
will serve to regulate the above expression.
Note that the factors of ${1 \over {H \pm i}}$
combine to have no effect in the corresponding classical
expression and that the $\theta$-functions in $Q^0 - \tau'$
are classically irrelevant for $\tau > \tau'$.  Thus, the classical objects
$[b]_{q^0 = \tau}$ and $\widetilde{[b]}{q^0 = \tau}$ differ only by
the factor $1/f(h_0)^2$.  It follows that as $B$ ranges over all
bounded operators built from the $Q(t)$'s and $P(t)$'s, $[b]_{q=\tau}$
ranges over an overcomplete set of functions on the space of solutions.

The usual calculation shows that convergence of the action of our operator on
some given state rests in the behavior of the matrix
elements
\begin{eqnarray}
( |E^*_0, E^*_1 &=& E^*_0, k^*\rangle, \widetilde{[B]}^L_{Q^0 = \tau}
|E_0, E_1 = E_0, k
\rangle)_{phys}  = 2 \pi f(E_0^*) f(E_0) \cr
&\times& \langle E_0^*, E_1^* = E^*_0, k^*|
\theta(Q^0 - \tau') B \theta(Q^0 - \tau') {H \over {H-i}}
\theta(Q^0- \tau)|E_0, E_1 = E_0, k \rangle
\end{eqnarray}
When the spectrum of $H_1$ is discrete,
$\theta(Q^0 - \tau)|E_0, E_1 = E_1, a\rangle$
is some normalizable state $|\psi(E_0, a, \tau)\rangle$ in ${\cal H}$
with
norm $z_{\tau}(E_0)$ which is a monotonically decreasing function
of $\tau$ and does not depend on $a$.
We will use this function to choose an appropriate $f$.
The case where $H_1$ has continuous spectrum can be dealt with in
much the same way, using wave packets and the factorization ${\cal H}
= {\cal H}_0 \otimes {\cal H}_1$, but we will not do so explicitly.

For the discrete case, fix $\tau' < \tau$ and let $n(E_1)$
be an enumeration of the energy levels of $H_1$ such that if $n = n(E_1)$,
the $n$-th energy level has eigenvalue $E_1$.  Similarly,
let $m(E_1)$ be the dimension, which we have assumed to be finite,
of the subspace of ${\cal H}_1$
with eigenvalue $E_1$.  Now, extend the functions $n(E_1)$ and $m(E_1)$
which until now were defined only on the spectrum of $H_1$ to all of ${\bf R}$,
say by defining them to be linear for $E_1$ not in the spectrum of $H_1$,
and let $f$ be any function such that
\begin{equation}
|f(E_0)| \leq {1 \over {n(E_0) \sqrt{m(E_0)} z_{\tau'}(E_0)}}.
\end{equation}
It then follows that
\begin{equation}
\label{april L bound}
|( |E^*_0, E^*_1 = E^*_0, k^*\rangle, \widetilde{[B]}^L_{Q^0 = \tau}
|E_0, E_1 = E_0, k
\rangle)_{phys} |^2 \leq {{4 \pi^2 ||B||^2 z^2_{\tau}(E_0)}
\over {n(E_0^*) n(E_0)
\sqrt{m(E_0) m(E^*_0)} z^2_{\tau'}(E_0) }}
\end{equation}
and
\begin{equation}
\label{april R bound}
|( |E^*_0, E^*_1 = E^*_0, k^*\rangle, \widetilde{[B]}^R_{Q^0 = \tau}
|E_0, E_1 = E_0, k
\rangle)_{phys} |^2 \leq {{4 \pi^2 ||B||^2 z^2_{\tau}(E^*_0)}
\over {n(E_0^*) n(E_0)
\sqrt{m(E_0) m(E^*_0)} z^2_{\tau'}(E^*_0) }}
\end{equation}
where $||B||$ is the operator norm of $B$ on ${\cal H}$.
Summing over $E^*_0$ and $k^*$, it follows that the norm of the
actions of $\widetilde{[B]}^L_{Q^0 = \tau}$
and $\widetilde{[B]}^R_{Q^0 = \tau}$ on  $|E_0, E_1 = E_0, k\rangle$
is bounded by
\begin{equation}
||\widetilde{[B]}^L | E_0, E_1 = E_0, k \rangle ||_{phys}^2
\leq {{4 \pi^2 ||B||^2 z^2_{\tau}(E_0)} \over {\sqrt{m(E_0)} n(E_0)}
z^2_{\tau'}(E_0)}
\sum_{n^*} {1 \over {(n^*)^2}}
\end{equation}
and
\begin{equation}
||\widetilde{[B]}^L | E_0, E_1 = E_0, k \rangle ||_{phys}^2
\leq {{4 \pi^2 ||B||^2 } \over {\sqrt{m(E_0)} n(E_0)}}
\sum_{n^*} {z^2_{\tau}(E^*) \over {z^2_{\tau'}(E^*)(n^*)^2}}
\end{equation}
where $n^*$ and $E^*$ are related by $n^* = n(E^*)$.
Thus,  we have
\begin{itemize}
\item{}
{\em Result 4} Let $B$ be a bounded operator on ${\cal H}_{aux}$
of a separable semi-bound system and let $B(t)$ be the associated
one parameter family of operators on ${\cal H}_{aux}$ defined through
\ref{Aev} with $B(0) = B$.  Then, for some function $f$ which
may depend on the system but is independent of the choice of
$B$,  the operator
\begin{equation}
\widetilde{[B]}_{Q^0 = \tau} = {1 \over 2}
(\widetilde{[B]}^L_{Q^0 = \tau} + \widetilde{[B]}^R_{Q^0 = \tau})
\end{equation}
given by \ref{bounded version} defines a {\em bounded} operator
with domain ${\cal H}_{phys}$.  Furthermore, if $B$ is
self-adjoint, then so is $\widetilde{[B]}_{Q^0 = \tau}$.
\end{itemize}
In particular, $B$ may be
of the form $B = (A {\rm sign}(P_0) + {\rm sign}(P_0)A) $
for some bounded $A$ that
commutes with both $P_0$ and $Q_0$ or we may have $B = A$ for such an
$A$.  It follows that we
have constructed a set of bounded operators $\widetilde{[B]}_{Q^0=\tau}$
on ${\cal H}_{phys}$
whose classical counterparts $[b]_{q^0 = \tau}$
form a complete set of functions on
the classical space of solutions (or on the classical
phase space) of the system.  We thus say that
\begin{itemize}
\item{}
The $\widetilde{[B]}_{Q^0=\tau}$ form a
complete set of bounded quantum observables.
\end{itemize}
Note that the bound in \ref{april L bound}
becomes zero as $\tau \rightarrow \infty$ since $z_{\tau}(E_0) \rightarrow 0$
and that the bound in
\ref{april R bound} also vanishes since the sum in \ref{april R bound}
converges absolutely and each term is a monotonically decreasing
positive function
of ${\tau}$.  Thus, we also have

\begin{itemize}
\item{}
{\em Result 5:} The $\widetilde{[B]}_{Q^0 = \tau}$ of Result 5
converge strongly to zero as $\tau \rightarrow \infty$.
\end{itemize}

\section{Discussion}
\label{diss}

We have seen that the construction of gauge invariants through integration
over manifolds can be combined with Hamiltonian methods for constrained
systems to
build a complete set of gauge invariant operators that are densely
defined in a physical Hilbert space
for a number of time reparametrization invariant models. A physical
inner product was identified and
this approach was shown to predict the collapse of spacetimes for
the appropriate Bianchi
cosmologies.  The primary method discussed
involved the use of an auxiliary Hilbert space
of the form $L^2({\cal Q})$, that is, containing functions that are
square  integrable over the configuration space ${\cal Q}$.

While the approach can be stated for an arbitrary finite dimensional
reparametrization invariant model, questions of convergence and
spectral analysis have only been addressed in narrower contexts.  The
most general rigorous results were obtained for already deparametrized
systems (\ref{ADS}), the free relativistic particle (\ref{RFP}), and
the separable semi-bound models of \ref{sep}.  Results
for other systems remain formal, but a forthcoming paper \cite{new}
will address the full Bianchi IX model.

Although only finite dimensional models were discussed in this
work, it might be hoped
that similar ideas could be applied to full quantum gravity.
However, a number of issues
would have to be faced.  These include the choice of representation (which may
be constrained by the parameterized dynamics of \ref{dyn}
in analogy with
unconstrained field theory \cite{Haag,GJ}), the choice of algebra
itself (see \cite{hisb,gp}), and the definition of the products of
operators that appear in our integrands (ultraviolet divergences).
On the other hand, another interesting extension would be to apply
these ideas at the $(C*)$ algebraic level without introducing a Hilbert
space at all.

Several interesting open problems also remain within
the current framework of finite dimensional models.
These include the search for
a general characterization of the convergence
of the integrals \ref{left} and \ref{right} and for the self-adjointness
(as opposed to just symmetry) of the resulting operators on ${\cal
H}_{phys}$.  It would also be useful to show that such operators form
a complete set in some inherently quantum mechanical sense (such as that
the resulting representation on ${\cal H}_{phys}$ is irreducible) and
to find a simple set of physically interesting operators whose
Hermiticity guarantees the uniqueness of the physical inner product.

Before concluding, a few remarks are in order about the limitations
of this approach and the delta-functions in \ref{sub pres}.  The point is that
while the technique of introducing an auxiliary Hilbert space of the form
$L^2({\cal Q})$ is
quite general and can (in principle) be applied to any
system which has a classical canonical description,
technical difficulties may prevent this method from being
{\it useful}.
The result may
be trivial in two ways.  First, it may be that when the Hamiltonian
constraint is chosen to be a Hermitian operator in the auxiliary
space its spectrum may not contain the value zero.  That is, there
may be no solutions to the constraint equation that are even
delta-function normalizable.  The other difficulty is that the
integral by which we would like to define our gauge invariants
may diverge on physical states.  We have seen that the matrix elements are
of the form $\langle E,k | [A]_{Q^0 = \tau}| E', k' \rangle =
\delta (E - E') f(E,E'k, k')$ for
eigenstates $|E,k\rangle$ of $H$ with eigenvalue
$E$ and some function $f$ so that, if
zero lies in
the point spectrum of $H$ and a solution to the constraint is
actually normalizable in the auxiliary space, the action of
$[A]_{Q^0 = \tau}$ on $|E, q^0 \rangle$ contains a divergent factor.
While it is possible that this divergence is regulated by the factor
$f$, this seems unlikely as we shall shortly see that this
divergence has a physical interpretation.

Note that both of these features
tend to occur when the spectrum of $H$ is discrete, such as in the
models \cite{ho}  where the constraint represents
a difference of two harmonic oscillator Hamiltonians or in
minisuperspace models that recollapse in a finite amount of
proper time $(\int N(t) dt)$.  This is the case for the LRS
Bianchi IX and Kantowski-Sachs models if the usual lapse
of general relativity
is used instead of our ``rescaled lapse."  A similar feature occurs in
the recollapsing models of \cite{GD}.
However,
discrete spectra are associated with
confining potentials as in \cite{ho,GD} and, if
the dynamics is complete, such systems
will generally have solutions for which
$q^0(t) = \tau$
at an infinite number of times in which case
$[a]_{q^0 = \tau}$
may diverge as well.  For such systems, one might
imagine replacing the integral over $t$ with
\begin{equation}
\lim_{t_{\pm} \rightarrow
\pm \infty} {1 \over {T(t_+) - T(t_-)}} \int_{t_-}^{t_+} dt
\end{equation}
and hope to obtain a finite result.  Equivalently, we may
take \ref{Omega action} as the {\it definition} of the operator $\Omega$
instead of \ref{form}.

If, however, in a gravitational model,
we take as one of our coordinates the parameter $\alpha = \ln (\det \ g)$,
then, because gravity classically admits gravitational collapse, we
know that
the ``potential" is not confining in the $\alpha \rightarrow
- \infty$ direction.  Thus, at least after a proper scaling of the
Hamiltonian constraint (in minisuperspace models) we expect that this
constraint
may be turned into a Hermitian operator for which zero lies in
the continuous spectrum of $H$ in
the auxiliary Hilbert space so that the methods of \ref{Dirac} will
be applicable.

\acknowledgements

The author would like to express his gratitude to Ranjeet Tate and Jorge
Pullin whose questions and comments prompted this analysis and to Bryce
DeWitt, whose influence manifests itself throughout this work.
Thanks also go to Jorma Louko for his helpful comments and
suggestions.
This work was partially supported by NSF grants
PHY93-96246 and PHY90-08502
and by research funds provided by The Pennsylvania State
University.

\end{document}